\documentclass[twocolumn,showpacs,preprintnumbers,superscriptaddress,amsmath,amssymb,floatfix]{revtex4-1}

\usepackage{graphicx}
\usepackage{dcolumn}
\usepackage{bm}
\usepackage{url}
\usepackage{amsmath}
\usepackage{verbatim}  
\usepackage{relsize}
\usepackage{textcomp}
\usepackage{booktabs}
\usepackage{xspace}  

\newcommand{\madx}{\textsc{mad-x}\xspace}

\newcommand{\bs}{$\beta^*$\xspace}

\newcommand{\sig}{$\sigma$\xspace}
\newcommand{\sign}{\sigma}

\newcommand{\mum}{$\mathrm{\mu}$m\xspace}

\newcommand{\fglob}{$f_\mathrm{glob}$\xspace}

\begin{document}


\title{Simulations and measurements of beam loss patterns\\at the CERN Large Hadron Collider}


 \author{R.~Bruce} \email{roderik.bruce@cern.ch} \affiliation{CERN, Geneva, Switzerland}
 \author{R.W.~Assmann}  \thanks{On leave from CERN, Geneva, Switzerland}  \affiliation{DESY, Hamburg, Germany}
 \author{V. Boccone} \affiliation{CERN, Geneva, Switzerland}
 \author{C. Bracco}  \affiliation{CERN, Geneva, Switzerland}
 \author{M. Brugger}  \affiliation{CERN, Geneva, Switzerland}
 \author{M. Cauchi}  \affiliation{CERN, Geneva, Switzerland}
 \author{F. Cerutti}  \affiliation{CERN, Geneva, Switzerland}
 \author{D. Deboy}  \affiliation{CERN, Geneva, Switzerland}
 \author{A. Ferrari}  \affiliation{CERN, Geneva, Switzerland}
 \author{L. Lari} \affiliation{CERN, Geneva, Switzerland} \affiliation{IFIC (CSIC-UV), Valencia, Spain}
 \author{A. Marsili}  \affiliation{CERN, Geneva, Switzerland}
 \author{A. Mereghetti}  \affiliation{CERN, Geneva, Switzerland}
 \author{D. Mirarchi}  \affiliation{CERN, Geneva, Switzerland}
 \author{E. Quaranta}  \affiliation{CERN, Geneva, Switzerland}
 \author{S. Redaelli}\affiliation{CERN, Geneva, Switzerland}
 \author{G. Robert-Demolaize}  \affiliation{BNL, Upton, New York, USA}
 \author{A. Rossi} \affiliation{CERN, Geneva, Switzerland}  
 \author{B. Salvachua} \affiliation{CERN, Geneva, Switzerland}
 \author{E. Skordis}  \affiliation{CERN, Geneva, Switzerland}
 \author{C. Tambasco}  \affiliation{CERN, Geneva, Switzerland}
 \author{G. Valentino}  \affiliation{CERN, Geneva, Switzerland}
 \author{T. Weiler}  \affiliation{CERN, Geneva, Switzerland}
 \author{V. Vlachoudis} \affiliation{CERN, Geneva, Switzerland}
 \author{D. Wollmann}\affiliation{CERN, Geneva, Switzerland}

\date{\today}

\begin{abstract}

The CERN Large Hadron Collider (LHC) is designed to collide proton beams of unprecedented energy, in order to extend the frontiers of high-energy particle physics. During the first very successful running period in 2010--2013, the LHC was routinely storing protons at 3.5--4~TeV with a total beam energy of up to 146 MJ, and even higher stored energies are foreseen in the future. This puts extraordinary demands on the control of beam losses. An un-controlled loss of even a tiny fraction of the beam could cause a superconducting magnet to undergo a transition into a normal-conducting state, or in the worst case cause material damage. Hence a multi-stage collimation system has been installed in order to safely intercept high-amplitude beam protons before they are lost elsewhere. To guarantee adequate protection from the collimators, a detailed theoretical understanding is needed. This article presents results of numerical simulations of the distribution of beam losses around the LHC that have leaked out of the collimation system. The studies include tracking of protons through the fields of more than 5000~magnets in the 27~km LHC ring over hundreds of revolutions, and Monte-Carlo simulations of particle-matter interactions both in collimators and machine elements being hit by escaping particles. The simulation results agree typically within a factor 2 with measurements of beam loss distributions from the previous LHC run. Considering the complex simulation, which must account for a very large number of unknown imperfections, and in view of the total losses around the ring spanning over 7~orders of magnitude, we consider this an excellent agreement. Our results give confidence in the simulation tools, which are used also for the design of future accelerators.

\end{abstract}

\maketitle

\section{Introduction}

 \label{sec:Intro}
The Large Hadron Collider (LHC)~\cite{lhcdesignV1,lhcJINST} at CERN is designed to collide protons with an unprecedented energy of 7~TeV and a total stored energy of about 362~MJ per beam. 
The operation started at a lower-than-design-energy of 3.5~TeV in 2010 and 2011, and in 2012 the energy was raised to 4~TeV, with the goal to reach the design parameters in the future. So far, a maximum of 146~MJ has been stored per beam during physics operation. The design stored energy of the LHC beams is at least a factor of~100 higher than in other hadron machines with superconducting magnets (HERA, TEVATRON, RHIC). 
 
Because of the high stored energy, the LHC beams are highly destructive. If protons deviate from the wanted trajectory so much that they hit the inside of the vacuum chamber, the induced heating can cause quenches (a transition to a normalconducting state) of the superconducting magnets that guide the beam around the ring, and possibly material damage. Even a local beam loss of a tiny fraction of a few $10^{-9}$ of the full beam (order of $10^6$ protons) in a magnet could cause a quench. Quenches must be avoided by all means during collider operation, since the recovery is a lengthy process that reduces the available time for collecting physics data. 

Therefore, all beam losses need to be tightly controlled. For this purpose, a multi-stage collimation system has been installed~\cite{jeanneret98,lhcdesignV1,assmann05chamonix,guillaume-thesis,assmann06,chiara-thesis,wollmann10}, in order to intercept unavoidable beam losses in a safe way. Unlike other high-energy colliders, where the main purpose of collimation is to reduce experimental background, the LHC requires collimation during all stages of operation to protect its elements. 

\begin{figure}[tb]
   \centering
   \includegraphics*[width=85mm]{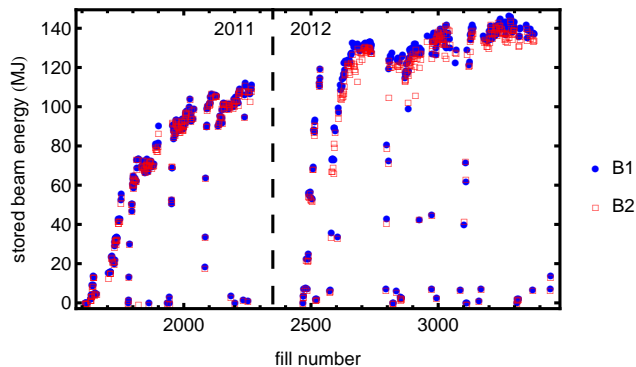}
   \caption{(color) Stored beam energy, for the two beams called B1 and B2, at the beginning of each LHC physics fill in 2011 and 2012 with proton collisions. The operational energy was 3.5~TeV in 2011 and 4~TeV in 2012. At the beginning of each year, a gradual ramp-up in intensity was performed for machine-protection reasons.}
   \label{fig:stored_energy}
\end{figure}

During the first LHC run 2010--2013 (called Run~I), the LHC collimation system has been very successful in protecting the cold magnets. No beam-induced quenches have occurred during physics operation with colliding beams, in spite of more than 100~MJ being routinely stored over many hours. The stored energy of the two counter-rotating beams, called B1 and B2, can be seen in Fig.~\ref{fig:stored_energy} for all physics fills in 2011 and 2012.

Even though the LHC collimation system has performed very well so far, the demands on collimation are increasing when the machine performance is pushed beyond the design values---e.g, the bunch intensity used in Run~I was above nominal, and with the number of bunches planned for the next LHC run, the intensity will be higher. Furthermore, future upgrades of the LHC, which are under study, foresee about a factor~2 higher stored beam energies~\cite{lucio_hllhc_ipac11}. 
  
Therefore, in order to ensure that future operation in the LHC will be as smooth and safe as during Run~I, it is vital to have a good theoretical understanding of the collimation system as well as the ability to predict local beam losses. For this purpose, we simulate the cleaning performance of the LHC collimators using the SixTrack code~\cite{schmidt94,robert05,nuria-thesis,jeanneret94-k2,claudia-thesis,claudia-paper}. SixTrack has previously been used at the design stage of the LHC to optimize the performance of the collimators~\cite{guillaume-thesis,chiara-thesis,lhcdesignV1}. SixTrack is still used to simulate present and future machine configurations. Therefore, in this article, we compare SixTrack results to measurements of LHC beam losses around the ring using beam-loss monitors (BLMs).

In order to perform a quantitative comparison with measurements, and to predict critical quantities such as the power density in the superconducting magnets, the proton losses produced by SixTrack are used as a starting distribution for a second stage of simulations of the secondary showers, induced by the lost protons. This is done using the Monte-Carlo code FLUKA~\cite{fluka1,fluka2008,vlachoudis09flair,ipac12_mereghetti_linebuilder}. At a few important locations, we investigate the quantitative accuracy of the full simulation chain.

First we give an overview of the LHC and its collimation system in Sec.~\ref{sec:lhccoll}, followed by a description of the simulation tools in Sec.~\ref{sec:sim_setup}, and of the BLM measurements in Sec.~\ref{sec:meas}. In Sec.~\ref{sec:2011_sixtrack_sim}, we present results of SixTrack simulations of the distribution beam losses in the ring. We study a machine configuration used in 2011 during Run~I at 3.5~TeV and compare qualitatively with BLM measurements, as well as analyze the sensitivity of the simulation result to imperfections and uncertainties in the starting distribution. In Sec.~\ref{sec:2011_FLUKA}, we present the FLUKA shower simulations of a few relevant regions and compare the results of the combined SixTrack and FLUKA calculation with BLM measurements. 

\section{The LHC and its collimation system}
\label{sec:lhccoll}

\begin{figure}[tb]
  \begin{centering}
  \includegraphics[width=8cm]{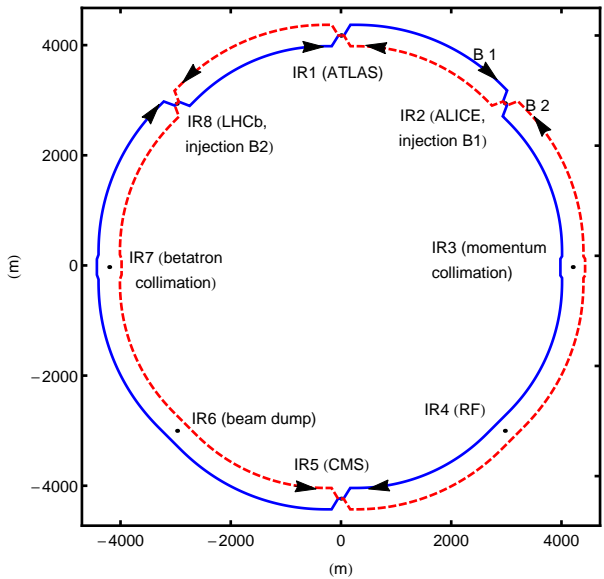}\\
  \end{centering}
  \caption{(color) The schematic layout of the LHC (the separation of the two rings is not to scale).
   The two beams collide at the four experiments ATLAS, ALICE, CMS and LHCb. Adapted from Ref.~\cite{bruce13_NIM_backgrounds}.}\label{fig:LHC-layout}
\end{figure}

The LHC is a 27~km synchrotron that consists of 8~straight sections, called insertion regions (IRs) and 8~arcs, as illustrated in Fig.~\ref{fig:LHC-layout}. Thousands of superconducting magnets, many operating at a temperature of 1.9~K, guide the two beams. Each IR houses either one of the four main LHC experiments (ATLAS~\cite{atlas}, CMS~\cite{cms}, ALICE~\cite{alice_Jinst}, and LHCb~\cite{lhcb_Jinst}), where the beams are brought into collision, or other equipment: the accelerating radio-frequency (RF) system is installed in IR4, the beam extraction takes place in IR6, and injection in IR2 and IR8. IR3 and IR7 are dedicated to the LHC collimation system. Some important parameters of the LHC---both the operational parameters in 2011 and 2012 and the nominal design values--- are summarized in Table~\ref{tab:machine_cond}. As can be seen, the design bunch intensity has been surpassed, and the achieved luminosity is almost as high as the design value in spite of a lower energy and fewer bunches. 

\begin{table} \centering
  \caption{Proton running conditions for physics in the LHC in 2011, 2012, and for nominal design parameters. The peak luminosity and \bs (the optical $\beta$-function at the collision point) refer to the high luminosity experiments ATLAS and CMS.}
  \label{tab:machine_cond}
  \begin{tabular}{lrrr}
   Parameter & 2011 & 2012 & Nom. \\ \hline \hline
   Beam energy (TeV) & 3.5 & 4 & 7 \\
  N. of  bunches & 1380 & 1380 & 2808 \\
  Average bunch intensity ($10^{11}$ p) & $1.2$ & $1.4$ & $1.15$  \\
  Peak stored energy (MJ) & 112 & 146 & 362 \\
  Horizontal and vertical \bs (m) & 1.5, 1.0 & 0.6 & 0.55 \\
  Peak luminosity ($10^{34}\mathrm{cm}^{-2}\mathrm{s}^{-1}$) & $0.35$ & $0.77$ & $1.0$ \\
  \hline
\end{tabular}
\end{table}


It is unavoidable that beam losses occur during collider operation. Apart from the wanted burn-off of protons in the collisions at the experiments, some colliding protons are scattered in elastic and diffractive events onto trajectories outside the machine acceptance, which happens also in collisions with rest gas. Furthermore, other  processes such as the long-range beam-beam effect~\cite{evans82}, intrabeam scattering~\cite{piwinski74,bjorken83}, and noise on the RF and orbit feedback systems cause a slow diffusion out of the beam core. More rapid losses can occur during changes of the machine configuration in the operational cycle or through beam instabilities. If losses occur in a superconducting magnet, the coils are heated by the induced hadronic and electromagnetic showers. If the temperature rise is high enough (about 2~K for the magnets operating at 1.9~K), a quench occurs. 

To avoid quenches and damage, the LHC has a system of about 4000~BLMs around the ring that detect losses during operation~\cite{holzer05,holzer08a} and trigger a beam dump if the losses are too high. BLMs are mounted on the outside of the cryostat of all quadrupoles in the LHC, as well as on all collimators and other elements that have been identified as potentially critical. The BLMs are ionization chambers, 50~cm long and filled with nitrogen. Since they are on the outside of the magnets, they intercept only secondary shower particles. 

A beam dump is triggered within 3~turns when a BLM detects a loss above a certain threshold. This delay is short enough to extract the beam before the magnetic field is significantly altered by a developing quench. The dump thresholds have been determined from the quench levels of the superconducting magnets, and from Monte Carlo simulations of the ratio between temperature rise in the coils and energy deposition in the BLM gas volume~\cite{report44,gschwendtner02,holzer08b}. 

In order to ensure stable running conditions, which are not interrupted by beam dumps, the continuously repopulated beam halo (the small fraction of particles surrounding a dense beam core) has to be safely removed by the LHC collimation system. The halo collimation is achieved by several stages, with the primary collimators, called TCP, closest to the beam, followed by secondary collimators (TCS) and active absorbers (TCLA), set at larger apertures. For optimal performance, the particles in the beam halo should first hit a TCP, and the TCSs should only intercept secondary halo particles that have been already scattered in, and escaped out from, upstream collimators. The TCP and TCSs, which are the closest collimators to the beam and hence intercept large beam losses, are made of a carbon fiber composite (CFC) to ensure high robustness. The TCLAs are meant to catch tertiary halo particles scattered out of the TCSs as well as showers from upstream collimators. The TCLAs are made of a tungsten alloy, in order to stop as much as possible of the incoming energy. On the other hand, they are not as robust as the CFC collimators and should therefore never intercept primary beam losses.

A three-stage system of this kind is installed both in IR7 and IR3, with the difference being that the horizontal dispersion in IR3 is much higher than in IR7. The IR3 collimators, which are usually more open than the IR7 ones, are thus used for momentum cleaning, while those in IR7 are used for betatron cleaning. The system in IR3 is built to intercept losses only in the horizontal plane, while the larger system in IR7 has a good coverage of the whole transverse space.

In addition to the dedicated insertions in IR7 and IR3, there are also collimators in most other IRs. Tertiary collimators (TCTs), made of a tungsten alloy, are installed in both beams about 150~m upstream of the collision points at all experiments [one TCT in the horizontal plane (TCTH) and one in the vertical (TCTV)]. 
They provide local protection of the quadrupole triplets in the final focusing system, which are the limiting cold apertures during physics operation. They are also important for decreasing the experimental background~\cite{bruce13_NIM_backgrounds}. 

Downstream of the high-luminosity experiments, ATLAS and CMS, there are special collimators to intercept the collision debris. Furthermore, at the beam extraction in IR6, special dump protection collimators are installed~\cite{lhcdesignV1} as a protection against mis-kicked beam in the case of extraction failures. Similarly, there are injection protection collimators in IR2 and IR8. 

\begin{figure}[tb]
  \includegraphics[width=8.5cm]{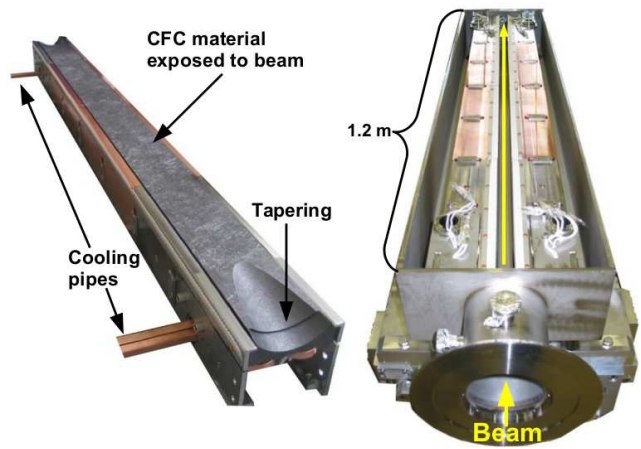}
  \caption{(color) A secondary collimator jaw made of CFC (left) and two parallel jaws installed in a collimator tank seen from the top (right), where the beam should pass in the center between the jaws. The primary collimators are similar but with an effective length of only 60~cm. }\label{fig:jaw}
\end{figure}

Most collimators consist of two movable jaws, with the beam passing in the center between them. A CFC collimator jaw is shown in Fig.~\ref{fig:jaw}. The collimator half-gaps are usually given in units of the local betatronic beam standard deviation 
\begin{equation}
\sign=\sqrt{\beta \epsilon/(\beta_\mathrm{rel}\gamma_\mathrm{rel})}. 
\end{equation}
Here $\beta$ is the nominal optical Twiss function, $\epsilon=3.5\,\mu$m the nominal normalized transverse emittance~\cite{lhcdesignV1}, and $\beta_\mathrm{rel}$ and $\gamma_\mathrm{rel}$ are the relativistic parameters. A collimator setting $n$ implies that the two jaws are positioned at a transverse distance of $\pm n\sign$ around the beam center. The collimator settings are kept constant 
from fill to fill but may vary in units of real beam standard deviations if the injected emittance is not nominal or the optics imperfect. These variations are, however, not relevant for the cleaning performance, since they do not alter the physics in how protons interact with the collimators and downstream magnetic elements. 

The collimator settings have been changed and optimized over the years as shown in Table~\ref{tab:coll_settings}. The calculated settings, of which a detailed explanation is beyond the scope of this article, are the result of an evolving optimization of the machine performance~\cite{bruce10evian,redaelli12_squeeze1m,assmann11tightMD,assmann11tightMD2,assmann11tightMD3,bruce11evian}. As an example, the TCP setting at 5.7~\sig\ in 2011 corresponds to half gaps of 1.5--2.2~mm.	

\begin{table} \centering
  \caption{Collimator half gaps, in units of beam standard deviation $\sign$, used during the LHC physics operation in 2011, 2012. They are shown together with the nominal design parameters. The reference beam energy $E$ is shown for each set of settings and it should be noted that $\sign$ scales as $1/\sqrt{E}$. }
  \label{tab:coll_settings}
  \begin{tabular}{lrrr}
   Parameter & 2011 & 2012 & Nom. \\ \hline \hline
   Beam energy (TeV) & 3.5 & 4 & 7 \\ 
   TCP cut IR7 (\sig) & 5.7 & 4.3 & 6.0 \\
  TCS cut IR7 (\sig) & 8.5 & 6.3 & 7.0 \\
  TCLA cut IR7 (\sig) & 17.7 & 8.3 & 10.0 \\
  TCP cut IR3 (\sig) & 12.0 & 12.0 & 15.0 \\
  TCS cut IR3 (\sig) & 15.6 & 15.6 & 18.0 \\
  TCLA cut IR3 (\sig) & 17.6 & 17.6 & 20.0 \\
  TCT cut IR1, IR5 (\sig) & 11.8 & 9.0 & 8.3 \\
  \hline
\end{tabular}
\end{table}


In spite of a sophisticated design, a small number of protons, initially hitting the TCPs, are not absorbed by the cleaning system. Instead, they leave the cleaning insertion on a perturbed trajectory and are possibly lost on the downstream machine aperture. The effectiveness of the collimators depends on their transverse openings as well as on their longitudinal placement in terms of betatron phase advance and dispersion---the theory is explained in Refs.~\cite{seidel-thesis, jeanneret98}. The collimation performance is usually quantified in terms of the local cleaning inefficiency $\eta$, which is defined as the ratio of local losses $N_\mathrm{loc}$ over a distance $\Delta s$ to the total losses on collimators $N_\mathrm{tot}$:
\begin{equation}
 \label{eq:eta}
 \eta=\frac{N_\mathrm{loc}}{N_\mathrm{tot} \Delta s}
\end{equation}

Operationally, the collimators are centered around the closed orbit through a beam-based alignment~\cite{valentino12}. Since it would be too time-consuming to align the jaw tilts with the beam envelope, they are kept parallel. 
Before high-intensity beams are allowed in the machine, the cleaning performance is qualified. This is done by provoking controlled beam losses, with a low-intensity beam, and observing the resulting loss pattern on the BLMs around the ring~\cite{wollmann10,valentino12,salvachua12evian,salvachua13ipacCleaning}. The losses as a function of the $s$-coordinate around the ring are called a loss map. In 2011, beam losses were created by driving the beam onto the third order resonance, while in 2012 a white-noise excitation from the transverse damper was used for some configurations~\cite{ipac11_hoefle_adt_blowup}. These two methods produce similar loss maps~\cite{moensIpac13adtVStune}. 

The rather lengthy procedure of alignment and qualification, which requires special low-intensity fills, can typically take 0.5--1.5~days and is only performed one to a few times per year~\cite{valentino12}. During the periods in-between, operation relies on machine reproducibility and the collimators are driven back to the previously qualified positions in every fill. 

\section{Simulation setup}
\label{sec:sim_setup}

\subsection{SixTrack}
\label{sec:sixtrack}

To simulate the cleaning of the LHC collimation system we use SixTrack~\cite{schmidt94,robert05}. It is a multi-turn tracking code that accounts for the six-dimensional phase space in a symplectic manner. SixTrack does a thin-lens element-by-element tracking through the magnetic lattice, accounting for multipoles up to order~20. It was initially developed for dynamic aperture studies and to achieve a high numeric stability when tracking particles over a large number of turns. SixTrack takes as input a sequence of magnetic elements, which can be created by \madx~\cite{madx}.

When a particle enters a collimator, a built-in Monte Carlo code~\cite{nuria-thesis,jeanneret94-k2,claudia-thesis,claudia-paper} is used to simulate the particle-matter interaction. Multiple Coulomb scattering and ionization energy loss are accounted for, as well as several point-like processes: nuclear elastic scattering, nuclear inelastic scattering (where it is assumed that the proton disintegrates---single diffractive events, where the proton survives, are treated separately), single diffractive scattering, and Rutherford scattering. Recent updates of the scattering routine~\cite{claudia-thesis, claudia-paper} include, among others, updated proton-proton elastic cross sections from the LHC~\cite{antchev13} and updated values of the single-diffractive cross sections based on a parametrization of the renormalized pomeron flux exchange~\cite{goulianos95}.

A particle is considered lost either when it hits the aperture---the particle coordinates are checked against a detailed aperture model with 10~cm longitudinal precision---or if it interacts inelastically inside a collimator. The exception to this rule is single diffractive events, where the incident proton could survive and exit the collimator. These protons, which often have significant energy offsets, are tracked further. The simulation output contains coordinates of all loss locations.

Different ways of generating the starting distribution of the primary beam halo for the tracking are available---in this article, we use two methods that we call \textit{annular halo} and \textit{direct halo}. For both methods we usually simulate separately the cases where particles hit first the horizontal, vertical or skew TCP. The annular halo is generated at the start of the LHC in IR1. The matched phase space in the collimation plane is populated uniformly in a thin segment around the normalized betatron amplitude corresponding to the TCP half opening. The shape in phase space is thus a thin hollow ellipse. 

\begin{figure}[tb]
  \centering
  \includegraphics[width=4.2cm]{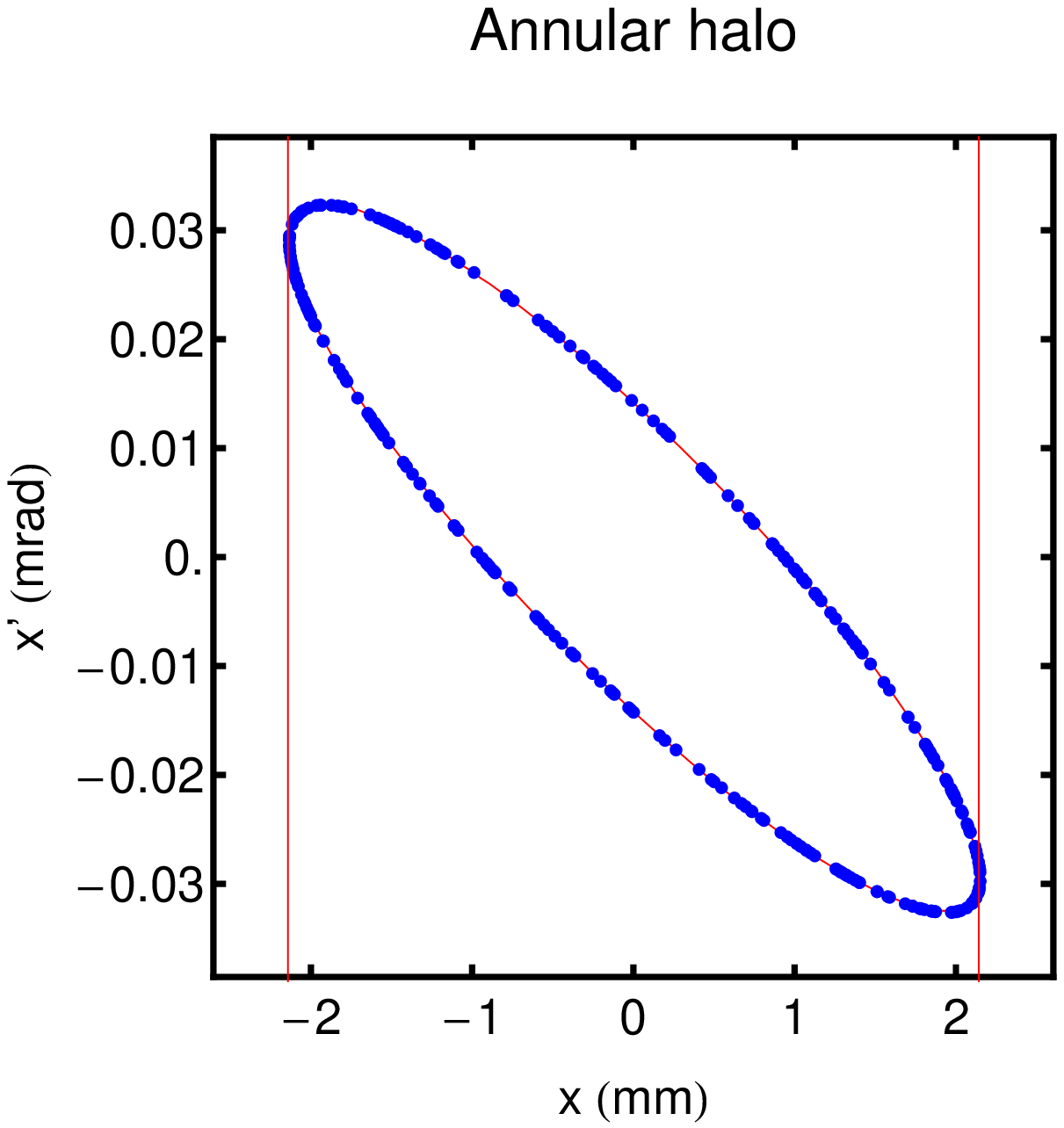}
  \includegraphics[width=4.2cm]{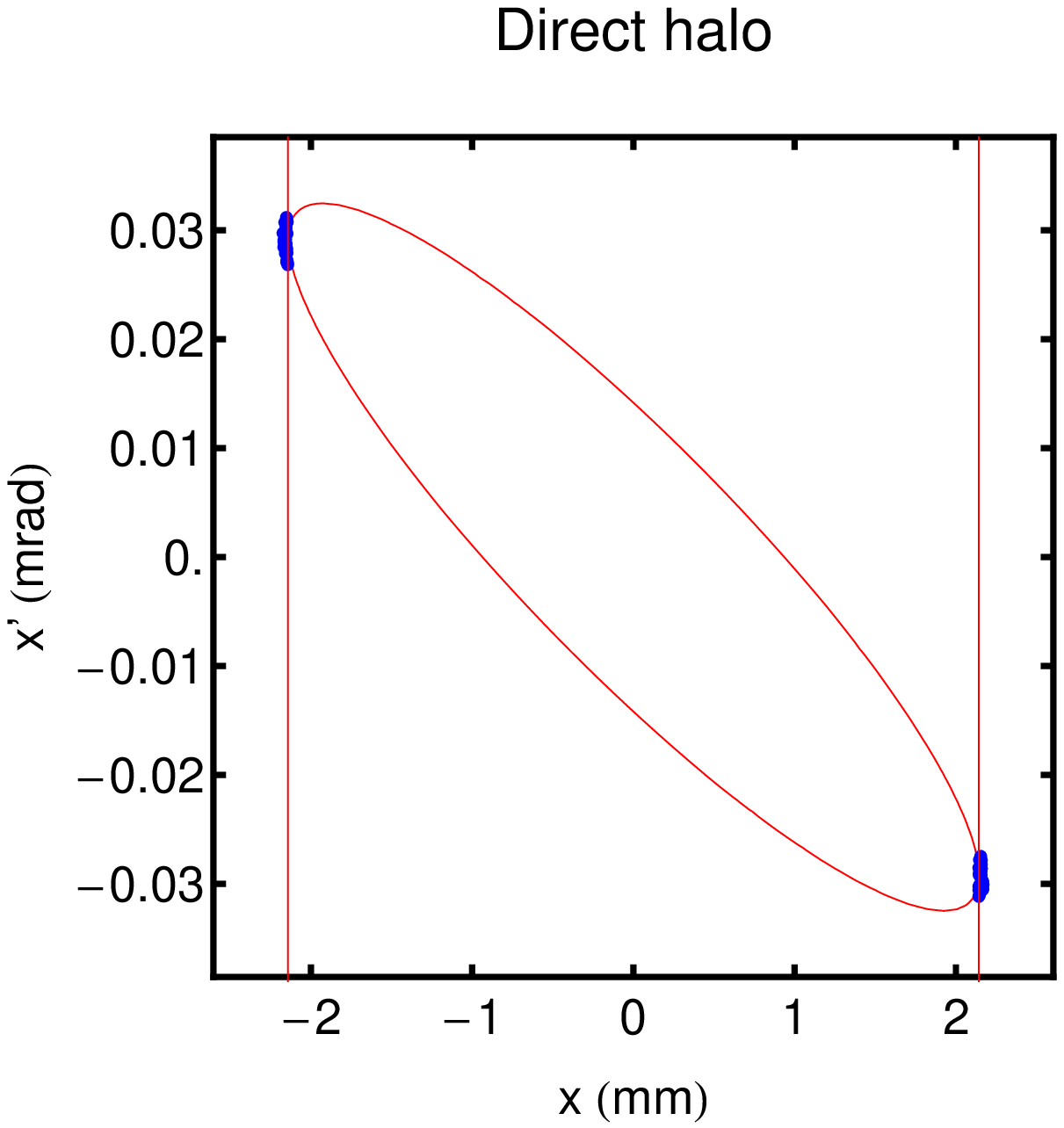}
\caption{(color) Examples of the horizontal phase space of the starting distributions at the TCP for annular halo (left) and direct halo (right). Each blue point represents a single particle, and the red ellipse the matched 5.7~\sig envelope. The vertical red lines represent the cuts of the TCP jaws.}
\label{fig:halo_phase_space}
\end{figure}

The direct halo is created directly at the collimator. It is identical to the annular halo except that particles in the collimation plane are generated only in the areas of the phase space that are outside the collimator cuts. Thus, with the direct halo, all halo particles hit the TCPs on the first turn, while with the annular halo, most particles circulate over many turns in the machine before they have the correct phase to hit the TCPs. Furthermore, when the initially well-defined annular halo shape travels from the start of the ring to the TCP, it is slightly deformed by non-linear magnetic fields, such as sextupoles, which could also alter the ratio of impacts between the two TCP jaws. The direct halo thus has the advantage that the impact distribution on the TCPs is much easier to control and that it is usually more efficient in terms of computing time, while the annular halo accounts for possible effects on the halo from non-linearities. For both the annular and direct halos, a 2D Gaussian distribution is assumed in the non-collimation plane. Example distributions of the phase space at the TCP for the two cases are shown in Fig.~\ref{fig:halo_phase_space}.

Neither of the two methods includes the diffusion that initially sends particles onto the collimators. This approach has the advantage that it becomes feasible to track many millions of halo particles to achieve sufficient statistics of losses, also at less exposed locations. If the beam core would be tracked as well, including diffusion, the needed computing time would rise by many orders of magnitude and the simulation would become impractically long. Instead, the starting coordinates of the tracking relies on assumptions on how the collimator intercepts the halo and therefore implicitly on the diffusion speed.

In order to achieve satisfactory statistics to resolve losses below the quench level, we usually track at least $6.4\times10^6$ protons for 200~turns---this is enough for the vast majority of the initial particles to be lost. Some of our simulations include $64\times10^6$ particles to resolve also smaller loss spikes.

\subsection{FLUKA}
\label{sec:fluka}

To simulate the BLM signals induced by proton losses, we use FLUKA~\cite{fluka1,fluka2008,vlachoudis09flair,ipac12_mereghetti_linebuilder}. FLUKA is a fully integrated particle physics Monte Carlo simulation code for the interaction and transport of particles and nuclei in matter. FLUKA is based on state-of-the-art models of physical interactions and tracks the initial particles as well as all created secondaries from the induced hadronic and electromagnetic cascades. The tracking is performed in a user-defined 3D geometry, including the detailed material composition and possibly magnetic fields. FLUKA has been developed over 25~years and is used in many different areas of nuclear science. A refined geometry of the LHC has been implemented over the last decade. It has been used to estimate energy deposition in the accelerator elements for various beam loss scenarios~\cite{cerutti14}, background to experiments~\cite{bruce13_NIM_backgrounds}, induced radioactivity~\cite{battistoni11} and radiation to electronics~\cite{roed12}.

\section{Measurements of LHC beam loss distributions}
\label{sec:meas}

\begin{figure}[tb]
  \centering
  \includegraphics[width=8.5cm]{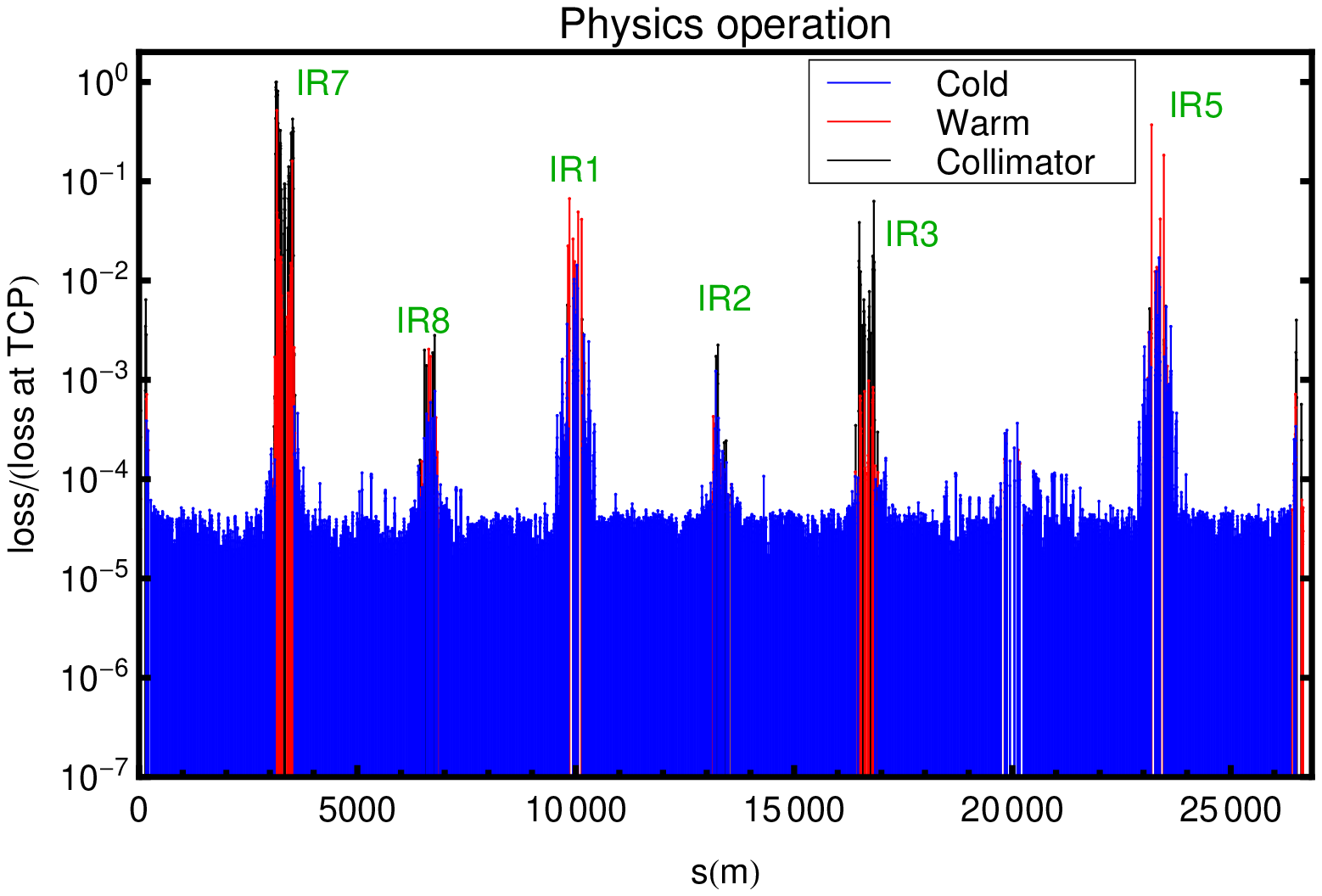}
  \includegraphics[width=8.5cm]{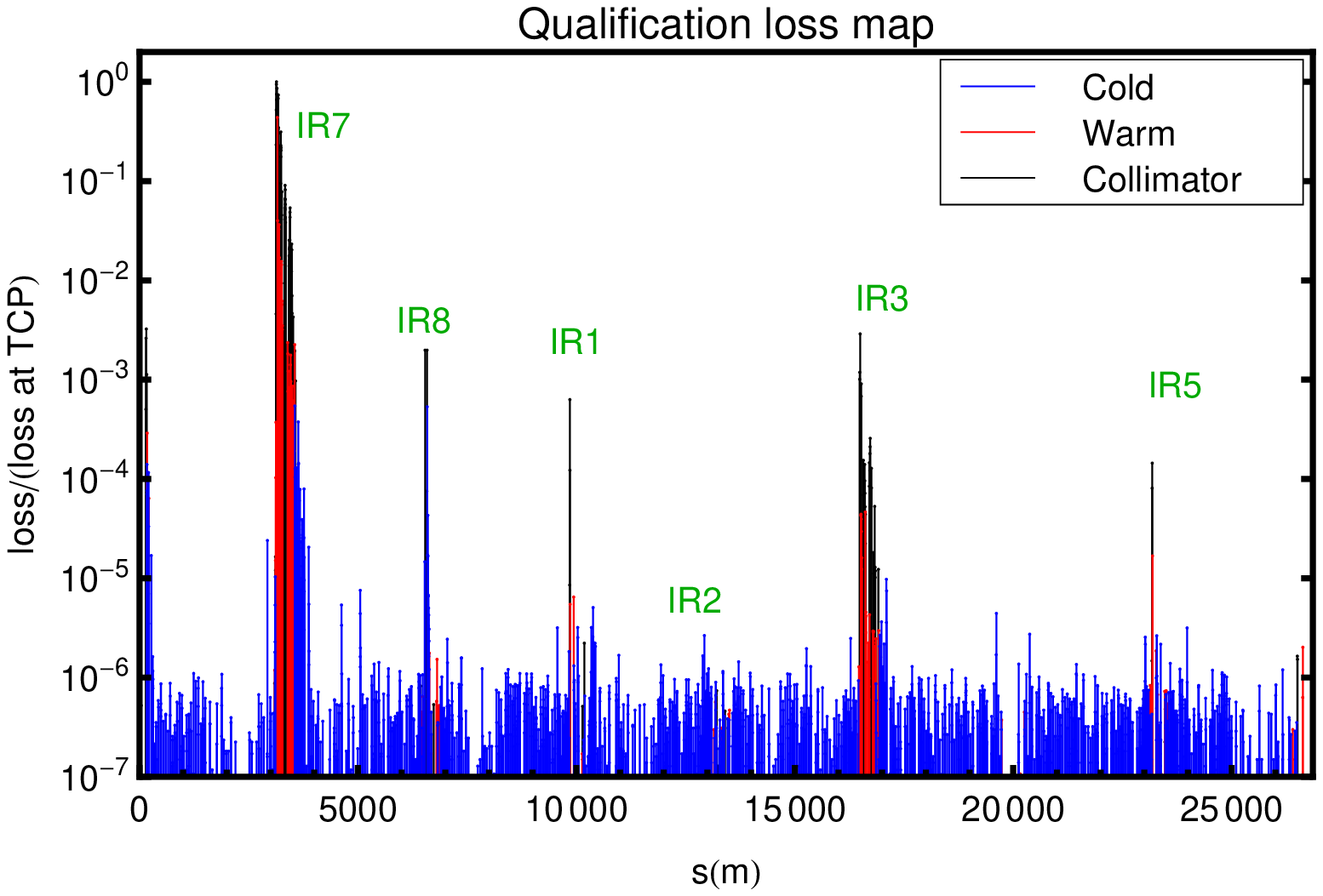}
\caption{(color) Examples of beam loss distribution around the LHC, measured with the BLMs using a 1.3~s integration time, from physics operation on August 17, 2011 (top) and from a qualification loss map of the betatron collimation system on March 11, 2011 (bottom). Here losses were provoked in B1 by crossing the third order resonance in the horizontal plane. The physics loss distribution was measured 10~minutes after the start of stable beams in fill number 2031.}
\label{fig:2011_meas_lossmaps}
\end{figure}

During LHC operation, the signals from all BLMs are continuously logged. Examples of the measured loss maps are shown in Fig~\ref{fig:2011_meas_lossmaps} with different colors for losses in cold or warm elements or on collimators. The top plot shows the losses during physics operation. The main loss locations are found on the collimators in IR7, but collision debris gives significant contributions around the high-luminosity experiments in IR1 and IR5. 

The bottom plot of Fig.~\ref{fig:2011_meas_lossmaps} shows instead a qualification loss map for the collimation system. Some significant qualitative differences can be observed, compared to the losses in physics: in this case, only one beam and one plane is excited, and the collisional losses at the experiments are negligible. Furthermore, since only the betatron amplitudes are excited, all primary losses occur in IR7, while during physics operation, there are also primary off-momentum losses in IR3. Another thing worth noting is that the signal-to-noise ratio is much higher during the qualification loss map, since the achieved absolute loss rate is higher. The only exception is if the physics beam suffers from high losses due to e.g. instabilities. 

For the sake of comparing SixTrack simulations to data later in this paper, it is therefore preferable to compare with the qualification loss maps, since we have increased resolution, and avoid the difficulties of disentangling the off-momentum and collisional losses from the betatron cleaning as well as the losses from the two beams, and have less uncertainties on the loss mechanism. Furthermore, we focus our comparisons on the betatron losses in IR7, since they are the limiting losses for the machine performance, and we do not study primary off-momentum losses in IR3 in this article.

\section{SixTrack simulations of the 2011 machine}
\label{sec:2011_sixtrack_sim}

\subsection{Perfect machine}
\label{sec:sixtrack_perfect}
\begin{figure}[tb]
  \centering
  \includegraphics[width=8.5cm]{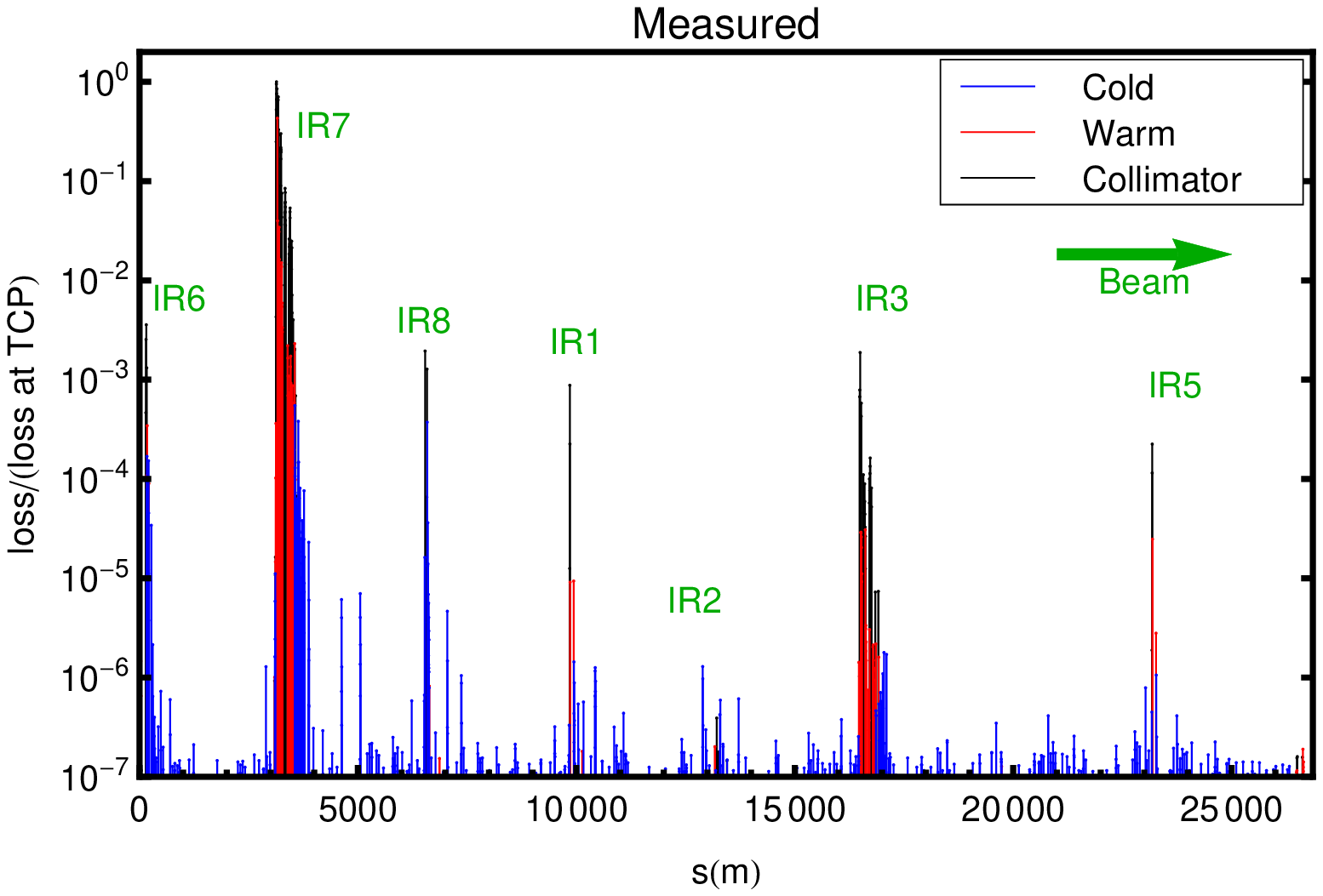}
\includegraphics[width=8.5cm]{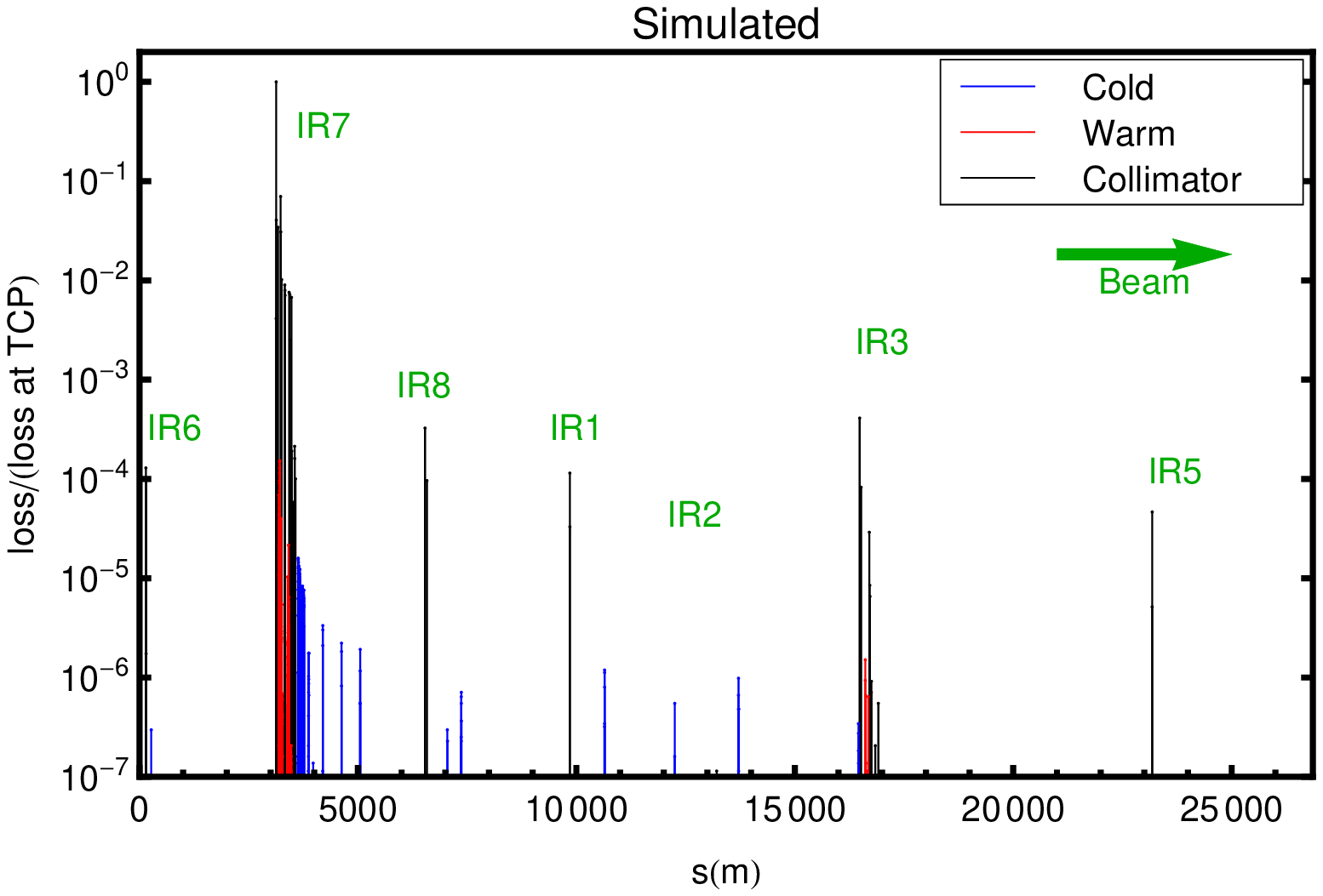}
\caption{(color) Beam loss distributions around the LHC as measured by BLMs during a qualification loss map on April 12, 2011, (top) and from a SixTrack simulation (bottom), with the results  binned in 1~m intervals. Both simulation and measurement assume a beam energy of 3.5~TeV and \bs=1.5~m. They are both normalized to the highest loss, and the initial losses occur in the horizontal plane in B1.}
\label{fig:lossmap_sixtrack_meas}
\end{figure}

The lower part of Fig.~\ref{fig:lossmap_sixtrack_meas} shows the simulated losses from SixTrack around the ring, for the case of a perfect machine without errors, using the 2011 machine configuration with \bs=1.5~m. The collimator settings are shown in Table~\ref{tab:coll_settings} and other machine parameters in Table~\ref{tab:machine_cond}. The starting distribution used in this example is an annular halo in the horizontal plane for B1 with an average impact parameter $b=13\;\mu$m on the TCP. Here $b$ is defined as the transverse depth into the jaw at which a particle is intercepted at its first hit. The influence of $b$ on the loss distribution is discussed in Sec.~\ref{sec:starting_dist}. The simulated losses are binned in 1~m intervals, except at the TCPs, which are considered as separate bins although they are only 60~cm long. 

The simulation result in Fig.~\ref{fig:lossmap_sixtrack_meas} is shown together with another example of a measured qualification loss map, taken at a different occasion than the one in Fig.~\ref{fig:2011_meas_lossmaps}. The observed loss pattern is very similar to Fig.~\ref{fig:2011_meas_lossmaps}, except that the background noise on the BLMs is less pronounced, since the achieved loss rate was stronger than in Fig.~\ref{fig:2011_meas_lossmaps}. 
Quantitatively, normalized losses above the background level can typically vary by a few tens of percent between measurements.

As can be seen in Fig.~\ref{fig:lossmap_sixtrack_meas}, there is a very good qualitative agreement between simulation and measurement. The main losses occur in IR7 at the collimators and the second most important loss location is IR3. The TCTs at the experiments are clearly visible as well as IR6, which also sees significant losses. We note that the simulation qualitatively predicts all potentially limiting cold loss locations. The result is representative also for vertical losses (similar loss pattern) and for losses in B2 (similar loss pattern, but with the beam going in the opposite direction). 

Significant quantitative deviations, where the simulation is by a few orders of magnitude lower than the measurement, are found at some locations, for example at the TCTs. However, the BLMs do not measure the direct proton losses shown for the simulation, but the secondary particles produced in the showers caused by the primary losses. The BLM signal per locally lost primary proton could vary significantly between loss locations, depending on the local geometry, materials, BLM location with respect to the loss position, and the spatial and angular distribution of the losses. Therefore, one cannot expect a high level of quantitative agreement when comparing the weighted convolution of all upstream showers in a BLM with the loss locations of primary beam protons. To do a quantitative comparison, it is therefore necessary to simulate also the showers. This is discussed in Sec.~\ref{sec:2011_FLUKA}.

\begin{figure}[tb]
  \centering
\includegraphics[width=8.5cm]{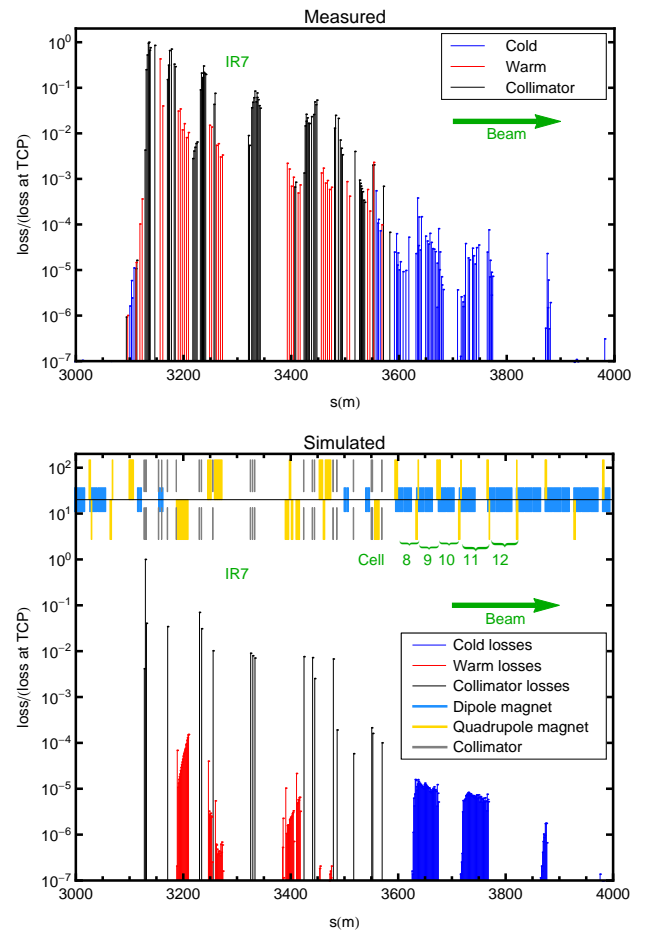}

\caption{(color) The same loss locations around the LHC from measurement (top) and SixTrack (bottom) as in Fig.~\ref{fig:lossmap_sixtrack_meas}, but zoomed in IR7. The layout of the main magnetic elements (quadrupoles and dipoles) as well as the collimators is also shown, together with the LHC cell numbers at the cold loss locations. }
\label{fig:lossmap_sixtrack_meas_zoomIR7}
\end{figure}

A zoom in IR7 of results in Fig.~\ref{fig:lossmap_sixtrack_meas} is presented in Fig.~\ref{fig:lossmap_sixtrack_meas_zoomIR7}. The highest losses occur, as expected, at the primary collimators and the loss levels decay along IR7. A small tail, a few orders of magnitude lower than the TCP loss, leaks to the cold magnets in the dispersion suppressor (DS) downstream of IR7. This location of the highest local cold loss in the ring is the limiting location for the LHC intensity reach from collimation cleaning~\cite{wollmann10}. 

Some qualitative differences can be observed: the measurement indicates a much denser loss pattern, with higher losses in the warm section, but also in the cold arc. This apparent discrepancy comes again from comparing the simulated number of lost protons with the measured BLM signals, which depend on the shower development. This is especially important for the warm BLMs, which are likely to intercept secondary shower particles created in upstream collimators. In the measurements in Fig.~\ref{fig:lossmap_sixtrack_meas_zoomIR7}, there seems also to be more collimators (black bars) than in the simulation. This is not the case---several BLMs are located at slots reserved for future collimators and are therefore displayed as such, although there is presently no collimator installed. These BLMs are also highly sensitive to the showers from neighboring collimators.

The simulated cold DS losses are grouped in two ``clusters'': the first one is centered around $s=3650$~m (see Fig.~\ref{fig:lossmap_sixtrack_meas_zoomIR7}) , and the second one is centered around $s=3740$~m. The average cleaning inefficiency is $\eta_{CL1}=8.6\times10^{-6}$~m$^{-1}$ in the first cluster and $\eta_{CL2}=5.2\times10^{-6}$~m$^{-1}$ in the second cluster, independently of the binning, while the  highest inefficiency in the cold parts of the ring, found in the first DS loss cluster, is $\eta_c\approx 1.9\times10^{-5}$~m$^{-1}$ with 1~m bins but goes up to $\eta_c\approx 5.2\times10^{-5}$~m$^{-1}$ with 10~cm bins due to steep aperture transitions. In total, the fraction of all simulated halo particles that are lost on other machine elements than collimators is \fglob=0.002. 

It should be noted that these $\eta$-values are calculated using Eq.~(\ref{eq:eta}), which is traditionally used for LHC collimation studies. It uses the total losses on all collimators as normalization and thus shows the local leakage out of the whole collimation system. The simulation result in Fig.~\ref{fig:lossmap_sixtrack_meas_zoomIR7} is, however, shown with a slightly different normalization using the losses on the TCP only. This allows a more consistent visual comparison with measurements but causes the $\eta$-values that can be read in Fig.~\ref{fig:lossmap_sixtrack_meas_zoomIR7} to be about 25\% higher than the numeric values quoted above.

The SixTrack simulation is based on an ideal optics and tune but the measurements were done while crossing the third order resonance, by adjustments of trim quadrupoles. Additional SixTrack simulations, where the fractional tune was changed to 1/3, show that this relatively small change in optics has no notable influence on the final loss distribution in most parts of the ring, except that losses decrease by about 20\% with the perturbed optics in the first part of the IR7 DS. The reason is an altered ratio of hits between the TCP jaws, caused by non-linearities acting in a different way on the annular halo. Further simulations, where only one of the TCP jaws is active at a time, show consistent variations in $\eta_{CL1}$. The real ratio between the jaws in the measurements is unknown, which introduces an uncertainty in the comparison.

The simulated losses are driven by qualitatively different processes at different locations around the ring. The simulated important cold losses in the IR7 DS are caused only by protons that have undergone single diffractive scattering---most of them come directly from the TCP. Their acquired scattering kick is not large enough for them to hit a TCS, TCLA or the aperture in the straight section, where the locally generated dispersion downstream of the TCP is low. However, the first bending magnets in the DS (see Fig.~\ref{fig:lossmap_sixtrack_meas_zoomIR7}), where the dispersion rises quickly, act as a spectrometer and over-bend the affected protons towards the aperture. The protons can circulate many turns and hit the TCP more than once before they undergo single diffractive scattering. The final loss in the DS occurs at the same turn as the scattering event.

\begin{figure}[tb]
  \centering
  \includegraphics[width=8.5cm]{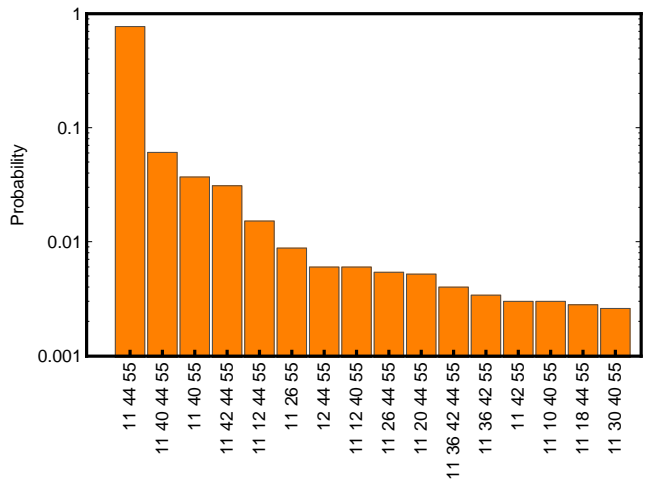}
\caption{The simulated distribution, at 3.5~TeV and \bs=1.5~m, of the most common histories of previously impacted collimators for the protons that are finally lost in a nuclear inelastic interaction on the TCTH in IR1 (55: TCTH IR1, 10--12: vertical, horizontal, and skew TCP in IR7, 16-44: different TCSs in IR7). For readability multiple entries of the same collimator have been neglected, e.g. particles with the history (11,11,44,55), where the two hits on collimator 11 occur on different turns, are counted in the same bin as (11,44,55). }
\label{fig:TCT_history_collimators}
\end{figure}

\begin{figure}[tb]
  \centering
  \includegraphics[width=8.5cm]{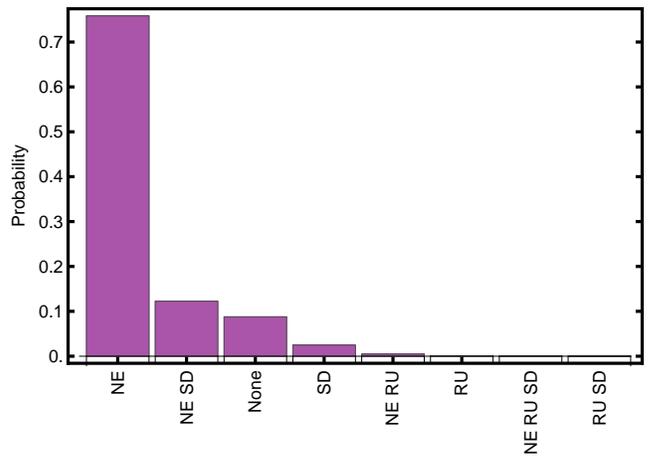}
\caption{The simulated distribution of histories of previously undergone point-like interactions in other collimators for the protons that are finally lost in a nuclear inelastic interaction on the TCTH in IR1 (NE=nuclear elastic scattering, SD=single-diffractive scattering, RU=Rutherford scattering). The particles which have not undergone any point-like interactions have acquired small offsets in angle and energy in repeated passages in collimators through multiple scattering and ionization. Multiple entries of the same physical processes have been grouped together, e.g. the left-most bin contains protons that have undergone nuclear elastic scattering one or several times but no other point-like interaction.}
\label{fig:TCT_history_interactions}
\end{figure}

Another important loss location is at the TCTs. Even though they absorb a lot of the impacting energy, a small fraction of the created shower particles leaks to the experimental detector, where they might cause unwanted signals. The losses at the TCTs should therefore be kept as low as possible.

The protons hitting the TCTs have previously hit both the TCP and one or several TCSs and in most cases made several turns after the first hit. An example of the distribution of histories of intercepted collimators, for particles lost on the TCTH in IR1, is shown in Fig.~\ref{fig:TCT_history_collimators}. All protons hit first the horizontal TCP in IR7---if instead a vertical excitation of the beam is considered, particles hit the vertical TCP first. One particular TCS is the source of more than 80\% of the TCT  losses in this example, however, the contributions from individual TCSs vary when the collimator settings or optics, and hence the phase advance, are changed.

The distribution of point-like physical interactions that the protons lost on TCTH in IR1 have undergone in upstream collimators is shown in Fig.~\ref{fig:TCT_history_interactions}. The distribution is similar at the other TCTs. In total, about 85\% of these protons have undergone nuclear elastic scattering and 10\% have in addition undergone single-diffractive scattering.


\subsection{Variations in starting distribution}
\label{sec:starting_dist}

Since our SixTrack studies start with an assumption on how the protons impact on the collimators, it is important that the initial conditions are as accurate as possible and that the effect of changing them is quantified. Early studies during the LHC design stage relied on theoretical results~\cite{assmann02} predicting $<b>\approx1\:\mu$m. With the machine in place, the LHC halo diffusion speed has now been measured using a collimator scan~\cite{valentino13_PRSTAB_diffusion}. These studies indicate $0.02\:\mu\mathrm{m}\lesssim b \lesssim 0.3\:\mu\mathrm{m}$ with single bunches during stable physics conditions, but $b$ is likely to be larger during beam instabilities and fast losses.

In the measurements in Fig~\ref{fig:lossmap_sixtrack_meas}, the beam was excited by crossing the third order resonance, which changes the diffusion speed compared to standard physics conditions, and hence $b$. The resonance crossing is very difficult to simulate accurately, since the halo dynamics at large amplitude depends strongly on unknown errors and non-linearities. We can, however, make approximate estimates. A SixTrack simulation has been performed with a different setup, where the initial distribution is Gaussian and includes the core and the optics has been matched to a fractional tune of 1/3. The result is rapid beam losses---after 1~s, only about 2\% of the beam remains, and the resulting average impact parameter is $<b>=15$~\mum. In reality, the loss process is much slower and continues over a few seconds. Therefore, the obtained $<b>$ is probably an overestimation, although the real $b$ is likely to be significantly larger than during physics operation. The discrepancy is likely to be caused by the slow approach to the resonance in the measurements, while the simulation starts directly at it. Furthermore, other non-linearities such as octupoles and the beam-beam effect introduce an additional tune spread which might alter loss rates. 

\begin{figure}[tb]
  \centering
\includegraphics[width=8.5cm]{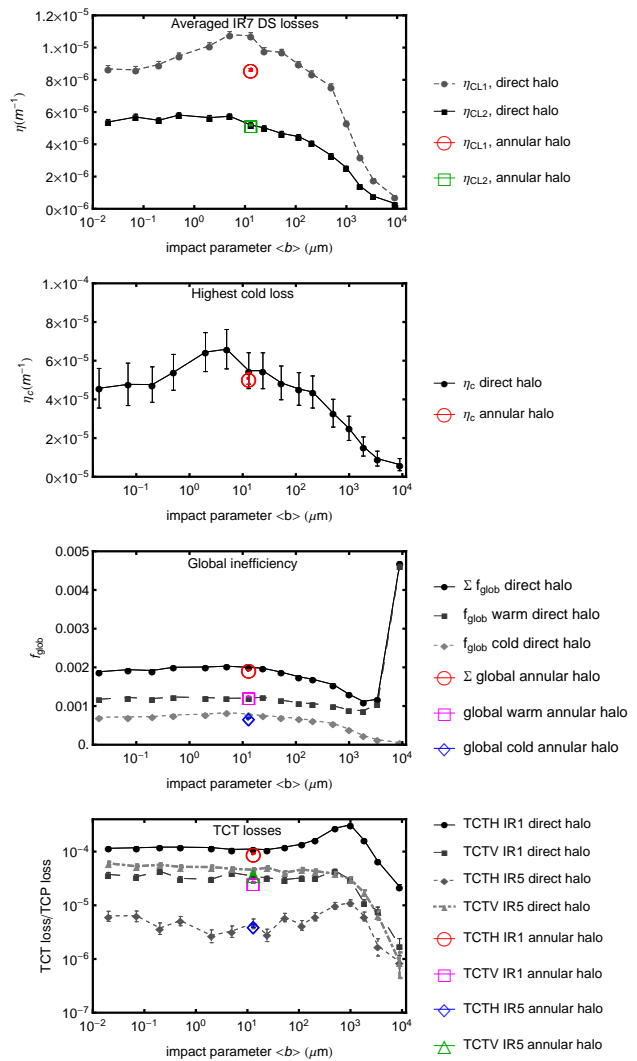}
\caption{(color) The dependence of various losses on the TCP impact parameter $b$, as simulated with SixTrack. 
All simulations were carried out for a horizontal halo in B1.}
\label{fig:impact_par_scan}
\end{figure}

Because of the uncertainty on $b$ and the differences between physics conditions and provoked losses, it is important to quantify the effect of the impact distribution on the simulated loss distribution. Therefore, we show in Fig.~\ref{fig:impact_par_scan} an example of the influence of $<b>$ on $\eta_{CL1}$, $\eta_{CL2}$, $\eta_c$, \fglob, and the leakage to the TCTs in the high-luminosity insertions. The scan in impact parameters has been performed using the direct halo, since $b$ is difficult to control precisely using the annular halo. 
As a comparison, we show, therefore, only one point from a simulation with annular halo in Fig.~\ref{fig:impact_par_scan}.

All losses are relatively independent of $b$, as long as $b$ is reasonably small, and decrease at large $b$. At small $b$, the impinging protons traverse only a very short distance inside the TCP, since they hit it with an inwards angle---the traversed distance is typically around 3~cm for an impact parameter of 1~\mum, while the nuclear interaction length is about 40~cm. Therefore, the protons are not likely to undergo a point-like interaction on their first TCP passage but instead they hit the TCP again at later turns and ``accumulate'' traversed length. 

As $b$ increases, protons are more likely to undergo a point-like interaction in the TCP on the first passage. At about $b=10$~\mum, the traversed distance in the TCP is close to a nuclear interaction length, and protons with $b\gtrsim 20$~\mum see the whole length of 60~cm. At larger $b$, the protons are more likely to be absorbed directly in the TCP, or by downstream collimators if they escape, and hence the general trend is that all losses outside IR7 decrease with increasing $b$.

Some exceptions to this can be observed: (i) since at very small $b$, protons usually do not undergo a point-like interaction on their first passage through the TCP, the non-linear fields deform the halo on the subsequent turns and alter the hit ratio between TCP jaws, just as for the annular halo. As $b$ increases, the jaw ratio goes towards 1 and therefore the two top plots of Fig.~\ref{fig:impact_par_scan} shows a slight increase of $\eta_{CL1}$ and $\eta_c$ up to $b\approx10$~\mum. (ii) At $b\gtrsim100$~\mum, the losses at the TCTH in IR1 increase (top curve of bottom plot of Fig.~\ref{fig:impact_par_scan}), since in this interval also the driving losses on the IR7 TCSs increase. 
At even larger $<b>$ above 1~mm, the TCT losses go to zero, since the protons are at such large betatron amplitudes that they are 
unlikely to leave IR7. (iii) At very large $b$, on the order of a few mm, the decreasing trend in \fglob is interrupted by a rapid increase in warm losses. This is caused by protons having so large amplitude that many of the escaping ones hit directly the warm aperture right downstream of the TCP. However, such large values of $<b>$ are extremely unlikely to occur in the machine.

It can be seen from Fig~\ref{fig:impact_par_scan} that, for the studied configuration, there is an excellent agreement between the simulations with annular and direct halo. The only exception is $\eta_{CL1}$ which shows a difference of 20\%. This difference is again caused by the ratio of hits between the two TCP jaws. 
The result from annular halo at $<b>=13$~\mum agrees with the direct halo when $b\rightarrow0$, where the jaw ratio is the same. For our purposes and compared to the overall span of the inefficiencies of many orders of magnitude, a 20\% difference is  very small.

From the analysis, we conclude that if $b\lesssim100\:\mu$m, $b$ does not play a significant role for the final loss distribution in our configuration. This means both that the qualification measurement can be considered as representative also for the loss distribution in physics, and that the uncertainty on $b$ is not important for the final result. At other energies, the threshold in $b$, above which the loss distribution changes significantly, is expected to show small variations as a function of the nuclear interaction length. Our results imply also that the computing time can be optimized by choosing a suitable larger $<b>$, where fewer simulation turns are needed.

\subsection{Machine Imperfections}
\label{sec:imperfections}

The previously shown simulation results assume a perfect machine. In reality, a number of unavoidable imperfections are present. A study of the influence of imperfections has been done in Ref.~\cite{chiara-thesis} for the case of 7~TeV and a nominal machine configuration. We apply here a similar, but extended, methodology to our 2011~example with appropriate adjustments of the parameters. 

\begin{figure}[tb]
  \centering
  \includegraphics[width=6.5cm]{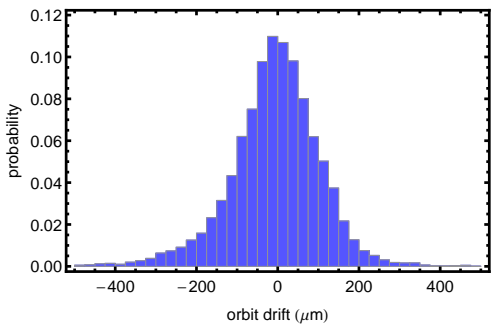}
\caption{The measured distribution of horizontal orbit drifts in B1 at the beam position monitors over six cells left and right of IR7. The orbit was sampled every minute during all physics fills in 2011.}
\label{fig:orbit_IR7}
\end{figure}

As the imperfections are unknown, our approach is to assign random imperfections according to some distribution and then run several seeds 
with different random errors, where each seed corresponds to one possible machine. Imperfections can be assigned either within SixTrack or in the input definition of the lattice. Several kinds of imperfections affect the result, coming both from the collimators and the rest of the machine: 

(i) The collimators are not always perfectly centered around the beam orbit, since the orbit drifts occur over time and between fills. Fig.~\ref{fig:orbit_IR7} shows an example of the measured orbit drifts at beam position monitors located in IR7 for B1 in the horizontal plane. Similar results are obtained for B2 and the other plane. The drifts are accounted for in SixTrack by adding random offsets to the centers of the collimators, using a standard deviation of 180~\mum, as found in the IR7 measurements. Uncertainties on the center coming from the collimator alignment are significantly smaller.

(ii) The tilt angle of the collimators with respect to the beam axis can suffer from angular misalignments of the collimator tank. We apply an rms tilt angle of 200~$\mu$rad in accordance with the studies in Ref.~\cite{chiara-thesis}.

(iii) Optics imperfections cause errors on the collimator gaps, errors on the phase advance between collimators, and a dispersion beating, which is enhanced by a imperfect closed orbit. Measurements~\cite{ipac11_vanbavinckhove,tomas12prstab} suggest an rms $\beta$-beating slightly above 4\% in 2011, which on average causes a gap error of about 0.17~$\sign$. Gap errors can be introduced either by random offsets within SixTrack, or through the lattice definition, where \madx is used to assign random magnetic errors and misalignments to achieve a realistic $\beta$-beating and orbit, which is partially corrected. The effects from phase advance and dispersion can only be included with the second method.

(iv) An imperfect jaw flatness can alter the effective length of material seen by impacting protons. Flatness measurements on some collimators are shown in Ref.~\cite{chiara-thesis}. We use the same approach as in Ref.~\cite{chiara-thesis} and apply a parabolic fit to the surface with a 60~\mum curvature for 60~cm long jaws and 100~\mum for 1~m jaws. This provides a result on the pessimistic side.

(v) The aperture of the magnets could be misaligned, which could alter the losses especially in the IR7 DS, where most losses occur on one side of the beam screen. We assume the design tolerances of misalignments for the different magnet types presented in Ref.~\cite{chiara-thesis} as a basis for the attribution of random errors.

\begin{figure}[tb]
  \centering
\includegraphics[width=8.cm]{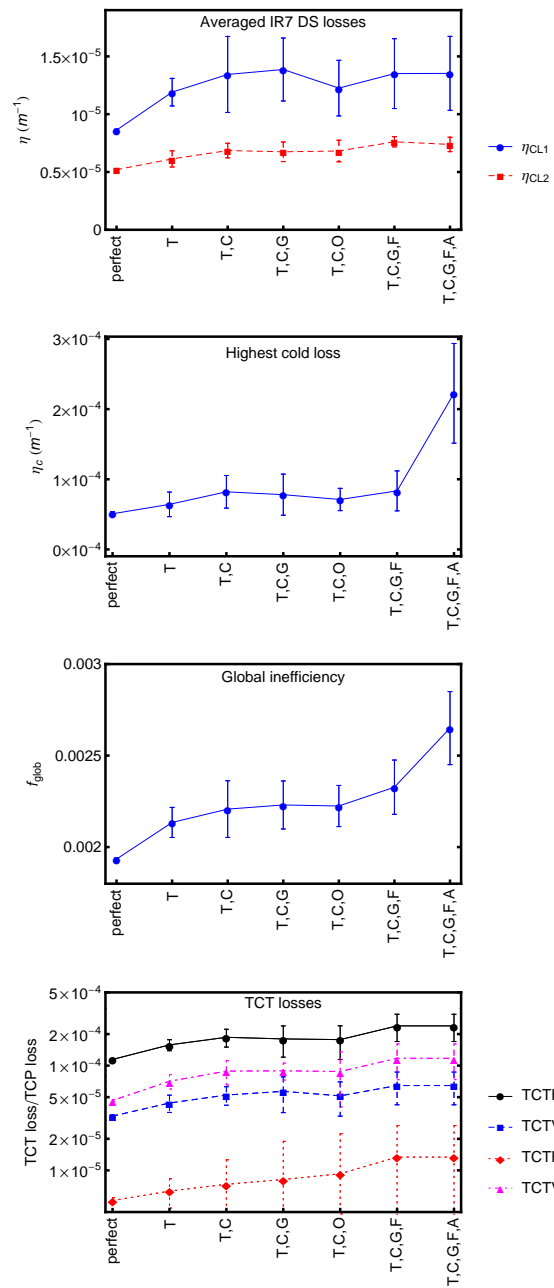}
\caption{The influence, as simulated with SixTrack, of different machine imperfections (T=tilt error, C=center error, G=gap error, O=optics errors introduced in the lattice in \madx, F=jaw flatness error, A=aperture misalignments) on different losses. }
\label{fig:imperfections}
\end{figure}

We summarize the influence of the different imperfections in Fig.~\ref{fig:imperfections}, which shows the simulated $\eta_{CL1}$, $\eta_{CL2}$, TCT losses, $\eta_c$, and \fglob for different configurations. All simulations were performed using a direct horizontal halo in B1 and $<b>\approx13$~\mum. It can be seen that when all imperfections are introduced, an increase of about 40\%--60\% is observed on $\eta_{CL1}$, $\eta_{CL2}$, and \fglob. The TCT losses increase by a factor 2--3, and $\eta_c$ by about a factor 4. 

The jaw tilts and flatness errors have similar effects, since they both shorten the distance traveled by protons inside the collimators. The gap errors and center errors change the normalized distances between the beam and the jaws. In the 2011 LHC configuration, the center errors have a stronger effect, and in some cases the gap errors are negligible. It should be noted that using an imperfect optics from \madx has about the same effect as the gap errors in a perfect optics, although the simulation with optics errors include imperfect dispersion and phase advances as well. These additional errors, therefore, are not important in our configuration. The importance of different types of imperfections varies between observables, e.g. the strongest effect on $\eta_c$ is caused by aperture misalignments, while at the TCTs, the tilt, center, and flatness errors dominate.

It should be noted that, in all cases, the error bars in Fig.~\ref{fig:imperfections} indicate the standard deviation between different imperfection seeds. The spread between different seeds increases significantly as more imperfections are added, which introduces higher uncertainties on a prediction of the real machine. With all imperfections, the spread is about 30\% for the more important loss locations. The statistical error on the mean values, not shown in the figure, is instead about 5\% in most cases, but higher at locations that intercept little losses. 

The computing time needed to achieve the results shown in Fig.~\ref{fig:imperfections} is very significant - each shown configuration corresponds to at least 20~seeds. Counting all seeds, a total of about $6\times10^{11}$ protons were tracked over 200--500 turns in the LHC. This simulation campaign would not be feasible without the use of a cluster where many simulations can be run in parallel.

\section{Quantitative comparisons with measurements using FLUKA}
\label{sec:2011_FLUKA}

Even though the inclusion of imperfections increases the losses outside IR7, and thus improves the agreement between simulation and data, we need to study the shower caused by the lost protons for a quantitative comparison. The inelastic interactions have to be simulated and the created secondary particles tracked in a detailed 3D geometry around the losses and out to the BLMs, accounting for the material composition and magnetic fields. For this study, we use FLUKA and restrict our study to BLMs only in some relevant regions. 

\subsection{Losses in the IR7 DS} 
We consider first the ten BLMs in the IR7 DS, which have the highest signal in measurements. They are all in cells~8 and~9 except one, which is in cell~11. These BLMs are very important since the losses in this region of the ring determine, together with the beam lifetime, a limit from collimation cleaning on the maximum allowed intensity. The FLUKA simulation is based on a model of IR7 up to the DS including the accelerator line and the tunnel, and incorporating magnetic field maps. The same model can be used to assess the power load in the superconducting magnet coils. Further details on this simulation setup are given in Refs.~\cite{bocconeReview11,ipac12_mereghetti_linebuilder,cerutti14}.

The starting conditions for FLUKA are positions of inelastic and single-diffractive interactions inside IR7 collimators, extracted from SixTrack. The FLUKA simulation starts with a forced interaction at these positions and tracks out-scattered particles to the DS. The shower in the DS is simulated separately in a second step to optimize the computing time. Possible showers from the warm section are not included. 
The simulation output is the energy deposition in the gas volumes of the BLMs.

The underlying SixTrack simulation, in B1 with a horizontal halo, was done for a perfect machine since it is very demanding in terms of computing time to simulate many imperfection seeds with FLUKA. To estimate the result for an imperfect machine, we scale up the FLUKA result for the perfect machine by the increase factor of nearby losses, as given by SixTrack, when imperfections are introduced. We calculate this factor over a 2~m interval upstream of the BLM. Clearly this is a simplification, since the shower is the convolution of all upstream losses. 
We include all imperfection sources mentioned in Sec.~\ref{sec:imperfections}. 

\begin{figure}[tb]
  \centering
  \includegraphics[width=8.5cm]{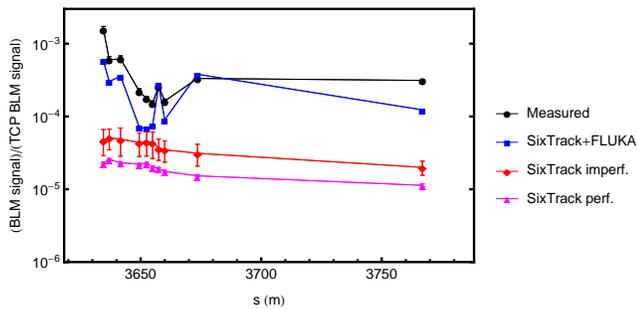}
\caption{The ratio of BLM signal, or particles lost, at the BLMs with the highest signal in the IR7 DS, to the signal at the horizontal TCP, in simulations and measurements for the 2011 machine for horizontal losses in B1. The errors on the SixTrack simulations indicate the standard deviation over different random seeds with collimator imperfections. The error bars on the measurements are taken as the standard deviation over the data set. }
\label{fig:ineff_IR7_DS}
\end{figure}

The FLUKA results, presented as the ratio of energy deposition in the BLM gas between the respective location and the horizontal TCP, are summarized in Fig.~\ref{fig:ineff_IR7_DS}. We show also the SixTrack output for a perfect and imperfect machine (average, with all error sources). 
For SixTrack, we divide the number of primary protons lost locally within 2~m upstream of the BLMs by the number of protons lost on the TCP. 
We show also the measured BLM signals, normalized to the BLM at the horizontal TCP and averaged over 5~different 2011 qualification loss maps in identical conditions but at different times over the year. 

Fig.~\ref{fig:ineff_IR7_DS} shows that the highest BLM signal in the cold part of the LHC occurs in cell~8 (the most critical location) and that the same maximum is reproduced in simulations. The general trend of the measurement in IR7 is very well reproduced by the combined SixTrack and FLUKA simulation, and the magnitudes of the measured BLM signals are typically underestimated by a factor~2. The largest discrepancy found is a factor 3. We consider this as an excellent agreement, considering all simulation uncertainties, including the unknown imperfections, and that the contribution of the shower from the long straight section is not included in the calculation. 

\subsection{TCT losses}
We study also the losses at the TCTs in ATLAS and CMS. The TCTs are important for experimental background~\cite{bruce13_NIM_backgrounds}. 
The tracking of protons after the final interaction that sends them onto a TCT involves large parts of the ring---the distance from IR7 to the TCTs is up to about 20~km. The protons lost at the TCTs have usually hit at least two other collimators before (see Fig.~\ref{fig:TCT_history_collimators}) and they often circulate many turns in the machine between these hits. Therefore, the tracking simulation of the TCT losses is significantly more complex and more sensitive to machine errors than the simulation of the IR7 DS, which relies mainly on a single-pass tracking from the TCP to the DS (about 500--700~m). 
On the other hand, the second simulation step with FLUKA is less demanding for the TCTs than for the IR7 DS, since a much smaller part of the geometry is involved. 

\begin{figure}[tb]
  \centering
  \includegraphics[width=8.5cm]{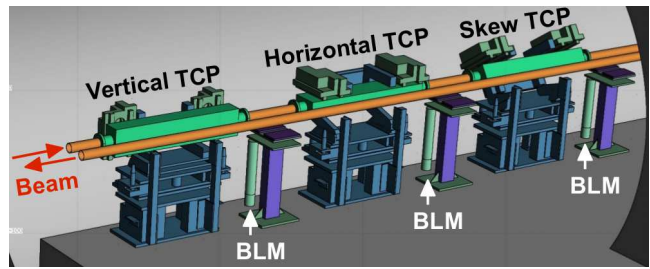}
\caption{(color) The 3D geometry surrounding the three IR7 TCPs, as implemented in FLUKA. The collimator jaws are contained in the green metal tanks and the BLMs are the upright cylinders below each collimator. }
\label{fig:tcp_geo}
\end{figure}

In order to estimate the relative BLM signals at the TCTs, we perform FLUKA simulations in two separate geometry models: one model of the three IR7 TCPs and one of the TCTH and TCTV (the layout is identical in IR1 and IR5). 
The FLUKA geometry of the TCPs is shown in Fig.~\ref{fig:tcp_geo}. As starting conditions we use the distribution of inelastic interactions in each collimator taken from the SixTrack simulation of the perfect machine. FLUKA simulations are done for both planes in B1. For each case, the TCTs in IR1 and IR5 are simulated in different runs. 

\begin{figure*}[tb]
  \centering
  \includegraphics[width=17cm]{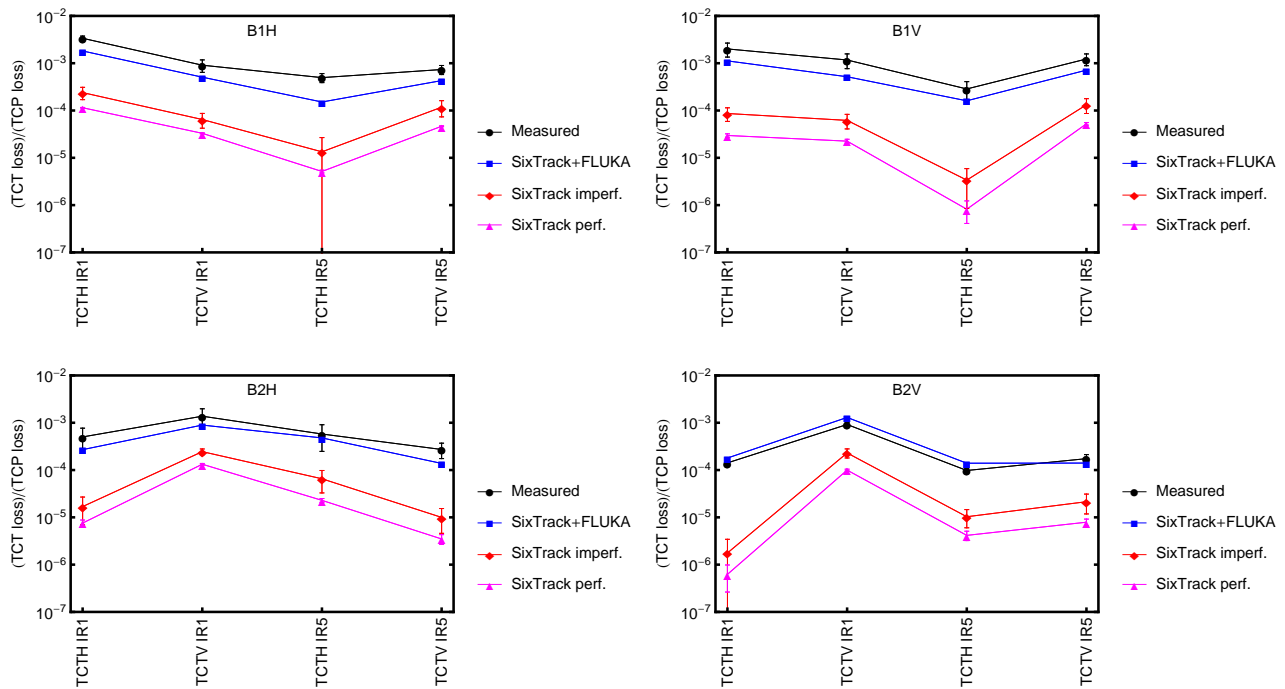}
\caption{The ratio of BLM signal, or particles lost, on horizontal and vertical TCTs to the TCPs in simulations and measurements in the 2011 machine. Simulation results are shown both from counting primary losses in SixTrack, as well as with a two-step simulation where FLUKA simulates the shower to the BLMs, starting from the SixTrack impacts in the simulations including imperfections. The errors on the SixTrack simulations indicate the standard deviation over different random seeds with imperfections.}
\label{fig:ineff_TCTs}
\end{figure*}

The output of the FLUKA simulations is BLM response matrices in IR7 and at the TCTs, which contain the energy deposition in each BLM per lost proton on each nearby collimator. The total energy deposition $S$ in a BLM is calculated as the sum over all nearby collimators of the number of SixTrack losses $N$ multiplied by the FLUKA response factors $R$. For example, $S$ at the horizontal ($h$) and vertical ($v$) TCTs is calculated as 
\begin{equation}
\label{eq:TCT_response}
\begin{pmatrix} 
 S_h\\
 S_v\\
\end{pmatrix}
 =
\begin{pmatrix} 
 R_{h\rightarrow h} & R_{v\rightarrow h} \\
 R_{h\rightarrow v} & R_{v\rightarrow v} \\
 \end{pmatrix}
 \begin{pmatrix} 
 N_h\\
 N_v\\
\end{pmatrix}
\end{equation}
where $R_{i\rightarrow j}$ is the response of the BLM associated to collimator $j$ to a loss on collimator $i$. The simulated quantity that we compare with the measurements is, as for IR7,  the ratio of the considered BLM signal, now at the TCTs, to the reference BLM signal of the most loaded TCP. 
The BLM response matrix includes the cross talk between the BLMs, which is especially important. Some TCTs intercept very few losses in SixTrack. However, the BLMs on these TCTs may still show significant signals caused by showers from the other nearby TCT. This causes in some cases an increase of the estimated BLM signal by more than two orders of magnitude compared to if only the losses at the collimator attached to the BLM are considered.

Table~\ref{tab:TCT_resp} shows the obtained $R$. The BLM response per lost proton is found to be up to about a factor~7 higher at the TCTs than at the TCPs. This is caused mainly by a difference in material (tungsten and CFC) and impact distributions, where the inelastic interactions occur much deeper in the TCTs (order of mm) than in the TCPs (tens of ~$\mu$m). Therefore, much more of the shower is developed within the TCT jaws than in the TCPs, and consequently more secondary particles reach the TCT BLMs. We observe a similar $R$ for the two planes in B1, in spite of small variations  in the impact distributions. Since these distributions are similar to B2, we assume the same $R$ in B2 as in B1. However, it should be noted that $R$ is likely to change if the collimator settings, beam optics, or energy is changed.

\begin{table} \centering
  \caption{TCT response matrices $R$ estimated with FLUKA for the 2011 physics run for B1 (3.5 TeV, \bs=1.5~m). The response is defined as the BLM signal per lost proton on the TCT, normalized by the response at the TCP. The total BLM signal is the sum of the contributions from both TCTs as shown in Eq.~(\ref{eq:TCT_response}). It should be noted that the presented values are likely to change if the optics, beam energy or collimator settings are modified.}
  \label{tab:TCT_resp}
\begin{tabular}{lrr|rr} 
  & \multicolumn{2}{c}{horizontal IR7 losses} & \multicolumn{2}{|c}{vertical IR7 losses} \\ \hline
  & BLM H  & BLM V & BLM H  & BLM V \\ \hline
TCTH IR1 & 6.9 & 1.1 & 7.2 & 1.1\\
TCTV IR1 & 0.4 & 3.3 & 0.4 & 3.3\\ \hline
TCTH IR5 & 6.3 & 1.2 & 6.7 & 1.5\\
TCTV IR5 & 0.4 & 3.2 & 0.5 & 3.1 \\ \hline
\end{tabular}
\end{table}

As it is not practically feasible to repeat the FLUKA simulation for all imperfection seeds, we assume the same $R$ in the imperfect machine as for the perfect one. The distribution of interactions inside the jaws does not change significantly---the main change is in the magnitude of losses. As for the IR7 simulation, we account for all imperfections when taking the average over all seeds. 

The simulated ratios of losses at TCTs to the TCP, both from SixTrack alone and including the FLUKA BLM response matrix, are shown in Fig.~\ref{fig:ineff_TCTs} for both beams and planes together with the measured average BLM ratios from the 2011 loss maps. 
In the SixTrack simulation of the perfect machine (magenta lines in Fig.~\ref{fig:ineff_TCTs}), the measured TCT leakage is underestimated by a factor 20--1000. Including imperfections increases the relative TCT losses by a factor 2--3, but does not change significantly the relative loss distribution between the different TCT. Including also the FLUKA BLM response matrix causes another increase by a factor 4--12 on the highest loss location and much more on the BLMs that are dominated by the shower from other collimators. As an example, the losses on the TCTH in IR1, during vertical B2 loss maps (left point of bottom right plot of Fig.~\ref{fig:ineff_TCTs}), increase from about 1\% of the measured value to 80\% when the FLUKA factor is accounted for. The inclusion of the showering changes the relative loss distribution between the TCTs to a shape that is similar to the measurements.

When comparing the combined simulation (blue lines) with the measurements (black lines), the loss distribution between the TCTs is very well reproduced in all four cases. We find an average underestimation of the measurements by a factor~1.6, but the discrepancy is never worse than about a factor~3 as for the IR7 DS. The vertical losses in B2 show a better agreement with a discrepancy of less than 30\%. We consider this an excellent result in view of the high complexity of the simulation chain and the many uncertainties.

\section{Conclusions}
In order to provide enough luminosity for the particle physics experiments, the LHC accelerator has to store proton beams of unprecedented energy. These beams are highly destructive and risk quenching the superconducting magnets if not controlled properly. Therefore, the LHC has a collimation system installed that should safely intercept beam losses. To ensure proper protection, a thorough understanding is required of the dynamics of the protons intercepted and out-scattered by the collimators.

For this purpose we use the SixTrack code to track particles through the magnetic fields and the jaws of the collimators over many turns. Given an initial assumption on the halo, SixTrack produces a resulting distribution of beam losses around the ring. The initial conditions have been estimated through measurements and simulations but have nevertheless significant uncertainties. However, the SixTrack results show that up to an impact depth of about 100~\mum, a value that is very unlikely to occur in the LHC machine, the initial conditions have a negligible influence on the final loss distribution. 

Machine imperfections, such as errors on collimator positions and the jaw surface, as well as magnetic errors around the ring, decrease the efficiency of the collimation system. In the studied LHC configurations, they cause the highest peak loss in a superconducting magnet to increase by a factor 4 on average, although the averaged losses in magnets increase only by 40--60\%. 

The simulated distribution of proton losses around the ring shows a very good qualitative agreement with BLM signals during provoked losses in the studied 2011 configuration.  The BLMs intercept secondary shower particles and not directly the beam protons. Therefore, for a quantitative comparison with the LHC measurements, a second simulation step is needed to assess the shower development between the initial losses and the detectors. We have done this at a few selected locations through a Monte Carlo simulation with FLUKA. We found an average discrepancy between measurement and the combined simulation of about a factor~2, and never worse than about a factor~3. We consider this as an excellent agreement in view of the total losses around the ring spanning over 7~orders of magnitude, the complexity of the simulation chain, and the very large number of unknown imperfections. 
Our results are based on tracking in total about $1.5\times10^9$ protons through the fields of the more than 5000~LHC magnets over hundreds of turns. The tracking was followed by shower calculations, based on a detailed implementation in FLUKA of several hundred meters of the LHC and the transport of a wealth of radiation components from TeV energies down to sub-MeV cutoffs. In terms of computational time, the simulation campaign is challenging and would not be feasible without parallelization on a cluster.


Apart from demonstrating that a complex physical process, such as multi-turn beam losses in the sophisticated LHC machine, can be accurately simulated, the good agreement with measurements gives us confidence that the simulation tools can be used to reliably estimate the beam cleaning performance in other LHC configurations, as well as in other machines, and to conclude on whether the efficiency is sufficient to maximize the availability and performance. Further validations are anyway planned at higher energy during the next LHC run.

Our simulation setup is already in use to assess future LHC configurations. New challenges arise if the proton energy is increased to 7~TeV and the total stored beam energy to about 700~MJ, as foreseen in the HiLumi LHC project~\cite{lucio_hllhc_ipac11}. In order to make sure that the total beam intensity will not be limited by the collimation cleaning performance and that the collimators can be operated smoothly, several upgrades of the LHC collimation system are under study~\cite{hilumi13_5.3,hilumi13_5.4,bruce14ipac_DS_coll,lechner14ipac_DS_coll,marsili14ipac_DS_coll_errors,valentino14_BPM_buttons}. The final need for upgrades, such as additional collimators in the IR7 dispersion suppressors, depends on a number of different parameters, such as the achieved beam loss rates and the quench limit of the LHC magnets at higher fields, where significant uncertainties exist. These parameters will be analyzed in more detail in the next LHC run in order to finalize the upgrade strategy.


\section{Acknowledgments}
We would like to thank B. Auchmann, R. de Maria, A. Lechner and F. Schmidt for input and discussions, and the CERN OP team for the support during the loss maps.

%

\end{document}